\documentclass[%
 reprint,
 amsmath,amssymb,
 aps,
]{revtex4-2}

\usepackage{graphicx}
\usepackage{dcolumn}
\usepackage{bm}

\usepackage{xcolor}\usepackage[version=3]{mhchem}

\usepackage[normalem]{ulem}

\begin{document}

\preprint{}

\title{
Anisotropy analysis of bamboo and tooth using 4-angle polarization 
micro-spectroscopy}
\author{
Meguya Ryu$^{1,2}$, Hsin-Hui Huang$^3$, Jitraporn Vongsvivut$^4$, Soon Hock Ng$^3$, Irma Dumbryt\.{e}$^5$, Donatas Narbutis$^6$, Mangirdas Malinauskas$^7$, Saulius Juodkazis$^{3,7,8}$, Junko Morikawa$^{8,9}$}

\affiliation{$^1$~School of Materials and Chemical Technology, 
The Institute of Science Tokyo, 2-12-1, Ookayama, Meguro-ku, Tokyo 152-8550, Japan}
\affiliation{$^2$~National Metrology Institute of Japan (NMIJ), National Institute of Advanced Industrial Science and Technology (AIST), Tsukuba Central 3, 1-1-1 Umezono, Tsukuba 305-8563, Japan}
\affiliation{$^3$~Optical Sciences Centre and ARC Training Centre in Surface Engineering for Advanced Materials (SEAM), School of Science, Swinburne University of Technology, Hawthorn, Victoria 3122, Australia}
\affiliation{$^4$~Infrared Microspectroscopy (IRM) Beamline, ANSTO‒Australian Synchrotron, 800 Blackburn Road, Clayton, Victoria 3168, Australia}
\affiliation{$^5$~Institute of Dentistry, Faculty of Medicine, Vilnius University, Vilnius, Lithuania}
\affiliation{$^6$~Institute of Theoretical Physics and Astronomy, Faculty of Physics, Vilnius University, Saul\.{e}tekio 9, 10222, Vilnius, Lithuania}
\affiliation{$^7$~Laser Research Center, Faculty of Physics, Vilnius University, Vilnius, Lithuania}
\affiliation{$^8$~World Research Hub Initiative (WRHI), School of Materials and Chemical Technology, 
The Institute of Science Tokyo, 2-12-1, Ookayama, Meguro-ku, Tokyo 152-8550, Japan}
\affiliation{$^9$~CREST-JST and 
The Institute of Science Tokyo, Meguro-ku, Tokyo 152-8550, Japan}


\date{\today}

\begin{abstract}
To investigate the anisotropic properties of biomaterials, two distinct classes are considered: polymer-based (e.g., cellulose in plants) and crystalline-based (e.g., enamel in teeth), each demonstrating distinct structural and functional characteristics. Four-angle polarization (4-pol.) spectral mapping of sub-1~$\mu$m bamboo slices was carried out in the mid-IR spectral range (2.5-20~$\mu$m) to reveal the 3D organization of the chemical bonding of cellulose using the key characteristic absorption bands associated with C-O-C and C-N vibrational modes. The longitudinal and transverse microtome slices revealed a switch between the presence and absence of dichroism in parenchyma cell walls and vascular bundles. The cell wall showed continuous alignment of the C-O-C stretching vibrational mode (8.6~$\mu$m/1163~cm$^{-1}$) down to the pixel resolution of $\sim 4~\mu$m (the step size in imaging) in the transverse slice; the cell wall thickness is $\sim 1~\mu$m. Thin microtomed slices of a tooth were measured in transmission and reflection modes. The single-point reflection measurements, performed using two perpendicular orientations, revealed orientational anisotropy in the enamel, which was absent in the dentin region. High sub-diffraction limited lateral resolution was numerically validated using a simplified-model of a Gaussian beam reading out material pixels with a defined orientation of absorption. It is shown that the orientation of small $\sim\lambda/10\approx 1~\mu$m objects can be revealed using a focal spot of $\sim\lambda/NA\approx 20~\mu$m, defining the diffraction limit for the objective lens with a numerical aperture $NA\approx 0.5$.
 \end{abstract}

\maketitle


\section{Introduction}


The main component of bamboo is cellulose, which is organized into complex 3D hierarchical arrangements, defining its mechanical and thermal properties~\cite{bam}. The anisotropy of any material is linked to the orientation and alignment of building blocks down to the microscopic or molecular level. Orientation patterns in bamboo cellulose are an active field of research, and polarization-based techniques are implemented~\cite{need}. In the 0.2-1~THz spectral region, bamboo has a refractive index of $\sim 1.4$ and absorption coefficient of $\sim 10~$cm$^{-1}$, with strong birefringence $\Delta n \sim 0.05$ at the orientations along and across fibers~\cite{Ichikawa:23}. It has been used to create THz polarization optics, e.g., a $\lambda/4$-waveplate for the linear to circular polarization conversion at $\sim 300~\mu$m wavelength (1~THz)~\cite{Ichikawa:23}. Anisotropy in temperature diffusivity of nano-cellulose fibers doped starch–polyurethane nanocomposite films was directly measured, showing a twofold difference in-plane $\alpha_\parallel = 2.12\times 10^{-7}$~m$^2$/s and out-of-plane $\alpha_\perp = 1.13\times 10^{-7}$~m$^2$/s~\cite{20m738}. The determination of anisotropies in optical, mechanical, and thermal properties is interlinked; for example, the refractive index is related to mass density, hence the birefringence to its gradient. One of the simplest, non-contact modes of material characterization is optical polarization-based analysis, e.g. using four polarization angles, which has now matured to cover a 
wide range of wavelengths from UV to THz. Recently, volumetric imaging to determine the 3D orientation of cellular structures was proposed using polarization-resolved light-sheet photoluminescence~\cite{Talon}.

\begin{figure*}[tbh]
\centering\includegraphics[width=18cm]{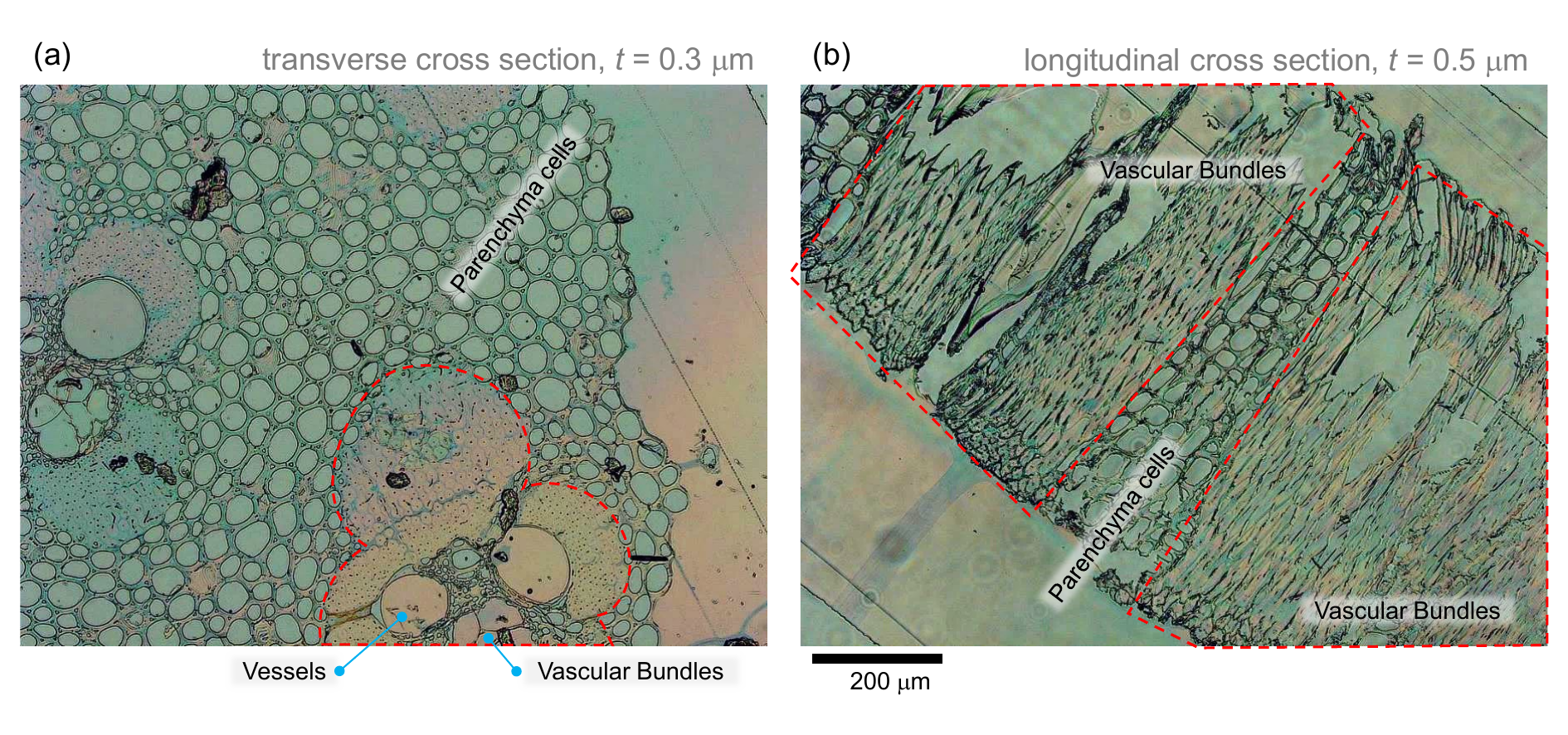}
  \caption{Optical images of the ultra-thin bamboo microtome (Leica EM UC6) sub-wavelength slices in transverse (a) and longitudinal (b) orientations. The bamboo sample was obtained from an ordinary chopstick. See the higher resolution images of the transverse cross-section in Fig.~\ref{f-hires}.} \label{f-bamb}
\end{figure*}

Tooth enamel has a complex microstructure that plays an important role in protecting the underlying dentin and pulp~\cite{imb2005nm,wegu15nm}. It is the most mineralized tissue in the human body, consisting of 95~wt.\% mineral (in mature enamel), 1~wt.\% organic matrix (proteins), and 4~wt.\% water~\cite{lit66aob}. On the microstructural level, the mineral (inorganic) component consists of crystal rods, also referred to as prisms, with each rod made up of bundles of nanometer-scale carbonated hydroxyapatite crystals, which are $\sim$~25~nm thick, $\sim$~100~nm wide, and $\geq$~100~nm long~\cite{bajd10jom,habs01aob}. The rods extend approximately perpendicular from the dentin-enamel junction to the tooth's surface~\cite{pars08jmsmm2317}. However, looking in more detail, the orientation of enamel rods varies depending on their location: in the outer enamel (closest to the tooth's surface), the rods are arranged in a nearly parallel manner, whereas in the region near the dentin-enamel junction, they form groups with oblique orientations~\cite{bajd10jom,dumi2025}. The distinctive microstructure of enamel, along with the spatial arrangement of its mineral and organic components, contributes to its mechanical properties, including high fracture toughness and resistance to crack propagation~\cite{bajd10jom,wegu15nm,imb2005nm,mirm2014nc}. 

Enamel exhibits all the characteristics of an anisotropic material, i.e. its mechanical properties such as hardness ($\sim$3 - 6~GPa), elastic modulus (70 - 120~GPa), and brittleness vary depending on the location, chemical composition, and arrangement patterns of the enamel rods~\cite{pars08jmsmm2317,cuyj02aob,habs01aob,mana06mos}. 
Studies have found that the hardness and elastic modulus of enamel increase with distance from the dentin-enamel junction: enamel near the surface of the tooth possesses the highest hardness and elastic modulus~\cite{pars08jmsmm2317}. In terms of absolute distance from the dentin-enamel junction, the highest gradient in these mechanical properties was observed within the cervical region~\cite{pars08jmsmm2317}. The chemical components and degree of mineralization, along with the location, play an important role in the mechanical properties of the enamel~\cite{cuyj02aob,braa07aob}. Both hardness and elastic modulus are positively correlated with calcium content~\cite{cuyj02aob,zhay14ijos}. When hypomineralized enamel was examined, significant reductions in hardness and elastic modulus were observed, with a relatively minor decrease in mineral content~\cite{mahe04b,mahe04ejos}. Calculations showed that a 3~GPa reduction in the elastic modulus could be expected with a 1\% reduction in the volume concentration of hydroxyapatite~\cite{stan81jms}. Finally, the mechanical properties of the enamel vary depending on the orientation of the rods, the arrangement of hydroxyapatite crystals within each rod, and their position along the length of the rod~\cite{pars08jmsmm2317,zhay14ijos,jeny11mbb}.

For teeth structure analysis at the micro- and even nano-scale, more sophisticated techniques such as 
synchrotron X-ray tomography~\cite{besc21mtc,besc23dj} can be employed. Some steps have already been taken in this direction, using synchrotron X-ray tomography to assess enamel demineralization in caries lesions in 3D~\cite{besc21mtc}, and to analyze the inner enamel during acid exposure~\cite{besc23dj}. A time-lapse imaging sequence was created by repeatedly scanning to detect changes in enamel structure due to acids~\cite{besc23dj}. Although enamel anisotropy plays a crucial role in the tooth's mechanical strength, its accurate assessment at the micro- and nanoscale remains challenging. Traditional mechanical testing methods often struggle to capture subtle changes in enamel properties, requiring advanced imaging and analysis techniques. A deeper understanding of enamel anisotropy is essential for improving dental treatment, developing biomimetic materials, and increasing knowledge of tooth biomechanics.

Another fundamental question in any optical analysis is how to improve resolution of the measurements in 3D space, especially at long IR spectral range discussed here. It is noteworthy that there is a fundamental link between the measurement of distance (space) and time (reciprocal to frequency)~\cite{1960}. Indeed, the height change $\Delta h$ (on  Earth) can be measured using an optical clock at frequency $\nu\propto 1/time$ (emission of the laser excited gas)~\cite{Katori1}. The frequency changes
(equivalent to the color/wavelength change) when 
the optical clock  raised to height $\Delta h$. The equivalent change in the optical transition frequency $\Delta\nu$ is $\frac{\Delta\nu}{\nu} =\frac{g\Delta h}{c^2}\approx 1.1\times 10^{-18}\times\Delta h\mathrm{[cm]}$. This constitutes the gravitational redshift and was experimentally verified by raising an optical clock by 1~m in the lab and measuring the red-shift (color/wavelength change). Later  the height of the Tokyo Skytree (450~m) was measured using the same optical clock at the ground level and at $\Delta h = 450$~m~\cite{Katori}. The upper clock ticks faster: at a 1~m height difference by 47~mHz~\cite{Katori}. Hence, by measuring frequency, a height difference (space) can be directly determined/measured. The frequency is probing local interactions, such as 
 the gravity in this case, and can be measured. In chemical maps, the frequency is measured and can be probed from nanoscale volumes with cross-sections of $\sim 10$~nm using a nano-sharp needle as shown for silk~\cite{19as3991}; the corresponding wavelength of the probed transitions would be 5-10~$\mu$m ($\Tilde{\nu} = 2000-1000$~cm$^{-1}$). Hence, it is no surprise that chemical mapping superior to the diffraction limit of the corresponding wavelength $\sim\lambda/NA$ can be achieved, as shown in this study.

Here, we show orientation maps of C-O-C and C-H cellulose IR absorption bands in the longitudinal and transverse microtome slices of bamboo (in respect to the growth direction) using the four polarization (4-pol.) method. Sub-wavelength $0.3-0.5~\mu$m slices were investigated at mid-IR $4000 - 500~$cm$^{-1}$ (2.5 - 20~$\mu$m wavelength, 120 - 15~THz) spectral window revealing 3D organization of cellulose structures: parenchyma cells and vascular bundles. Experiments were carried out on the IR micro-spectrometry beamline at the Australian Synchrotron (ANSTO). For comparison, a tabletop FTIR microscope was used to measure the reflectance spectra of the microtomed hard bio-tissue, the tooth. 
Anisotropy of reflectance was apparent in the enamel region. By using a simplified numerical model, it is shown that orientation can be determined from objects/structures with sub-diffraction-limited feature sizes in the IR chemical fingerprinting spectral window.

\begin{figure*}[tb]
\centering\includegraphics[width=18cm]{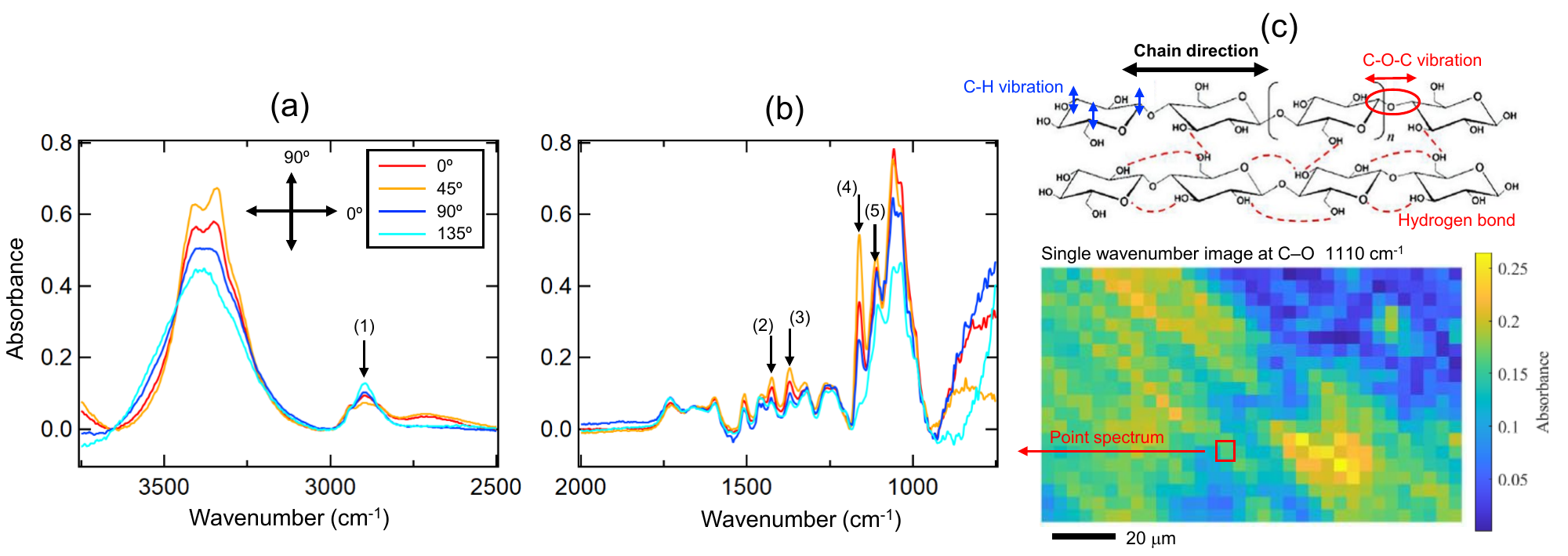}
  \caption{(a, b) Single point IR absorbance spectra at two spectral windows with marked bands (1-5) from the spectral map (c); inset in (c) shows cellulose structure (adapted from ref.~\cite{bam}; see Crystal Maker generated cellulose structure in Fig.~\ref{f-lsli}(b)).} \label{f-poin}
\end{figure*}
\begin{figure*}[tb]
\centering\includegraphics[width=18cm]{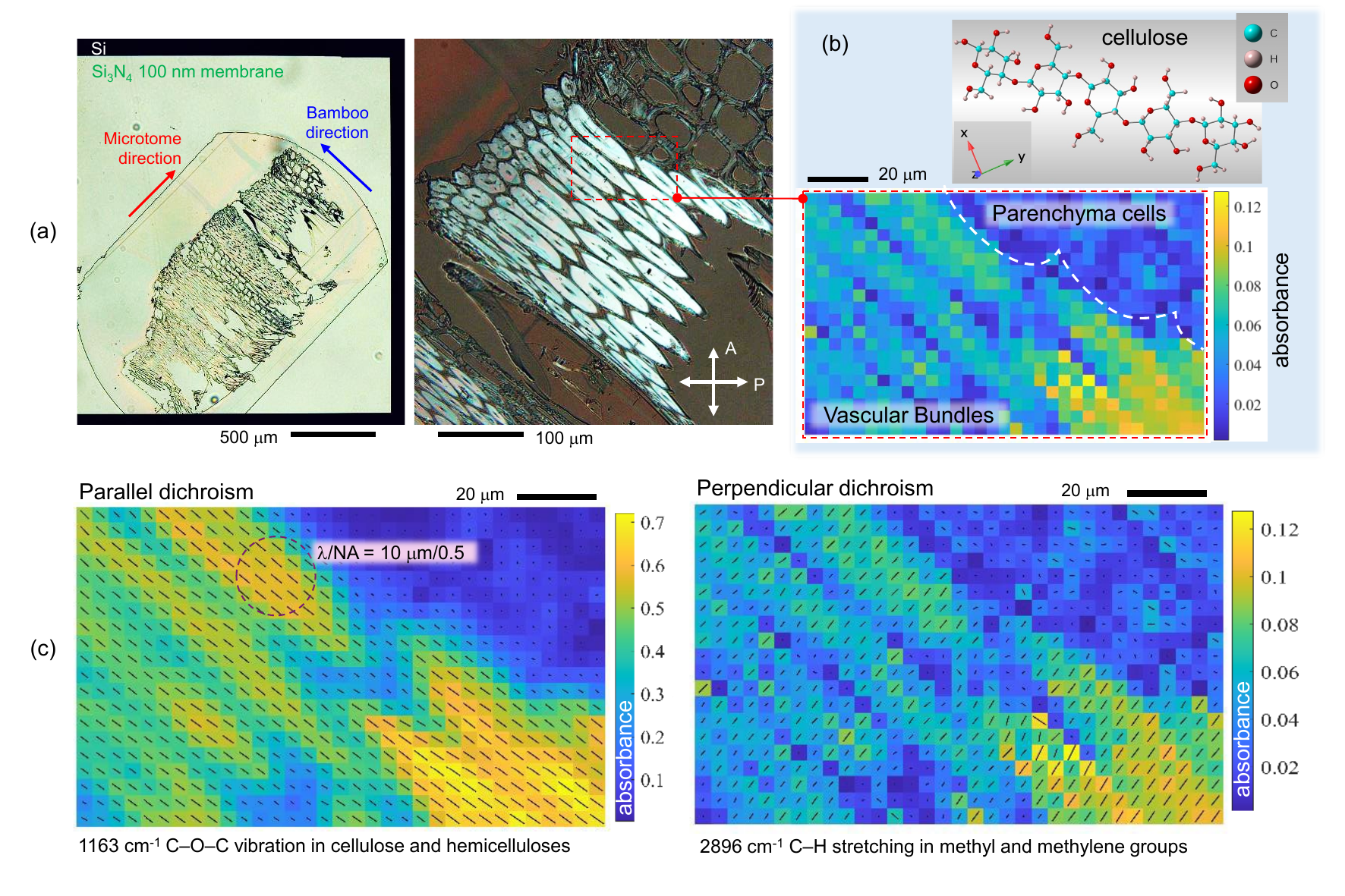}
  \caption{(a) Optical and polariscope views of bamboo microtome slice on a \ce{Si3N4} 100-nm-thick membrane. (b) Absorbance map at 2896~cm$^{-1}$ C-H stretching band; top inset: cellulose structure generated by CrystalMaker (Version 10.8.2). 
  (c) Absorbance map at 1163~cm$^{-1}$ C-O-C band. (d) Absorbance map at 2896~cm$^{-1}$ C-H band.} \label{f-lsli}
\end{figure*}

\section{Experimental: samples and methods}

\subsection{Microtome of bamboo}

Bamboo samples were taken from ordinary chopsticks. A peeled-off mm-sized flake of bamboo was embedded into a white light photo-curable resin (Aronix D-800, Toagosei Ltd., Tokyo, Japan). Ultra-thin microtome slices were taken along and across the direction of bamboo growth (evident by a filamentary pattern) using an ultra-thin microtome Leica EM UC6. Longitudinal and transverse microtome slices of 0.5 and 0.3~$\mu$m, respectively, were made and placed onto a 100-nm-thick \ce{Si3N4} membranes for imaging in the visible and IR spectral ranges. 
Figure~\ref{f-bamb} shows optical images of micro-slices with typical patterns of parenchyma cells and vascular bundles. Complex structure is apparent, with a clear shape anisotropy of vascular bundles, which are elongated along the plant's growth direction.

\subsection{Microtome of tooth}\label{MTooth}

Following extraction, a maxillary premolar sample was prepared according to the guidelines of the International Organization for Standardization (ISO/TS 11405; 2015)~\cite{ISO1140503}. The tooth selection criteria were as follows: (a) intact (i.e., healthy, undamaged) enamel with no white spots, signs of dental fluorosis, or enamel hypoplasia; (b) no pre-treatment with any chemical agents (such as hydrogen peroxide); (c) no previous dental treatment; (d) non-traumatic tooth extraction procedure. The tooth was stored in an individual tube inside water before slicing and microtome processing. Informed consent was obtained from the participant. The study protocol was approved by the Vilnius Regional Committee on Biomedical Research Ethics (approval number 2022/9-1458-933). All procedures were conducted in compliance with the relevant guidelines and regulations.

The tooth sample 
was removed from the water, air-dried for 24 hours, and then placed into acrylic resin molds. It was subsequently filled with a Struers epoxy resin mixture consisting of 150~g of EpoFix and 18~g of EpoFix hardener. The mold was placed under vacuum (CitoVac, Struers, Copenhagen, Denmark) for 30~minutes to remove excess bubbles. The sample was left to set at room temperature for 24~hours before being removed from the mold.
It was then sliced vertically into three sections to expose the enamel and dentin parts of the tooth. This was achieved using a 6-inch diamond Wafering Blade (WB-0065HC, PACE Technologies, Tucson, AZ, USA) on a Precision Sectioning Machine (IsoMet 1000, Buehler, Lake Bluff, IL, USA), with the speed set at approximately 150~$\pm$50 rotations per minute (rpm).
To achieve a mirror polish suitable for IR measurements, an automatic grinding and polishing machine (Tegramin-25, Struers, Copenhagen, Denmark) was used. The polishing protocol included:
1) grinding with finer silicon carbide papers ($\#$2000, Struers, Copenhagen, Denmark; Plate: MD-Gekko, Struers, Copenhagen, Denmark; speed: 150 rpm; applied pressure: 15~N) for 1 minute per side, while pausing the machine every 30 seconds to ensure the dentin or enamel were not over-polished.
Then, 2) polishing with water-based diamond suspensions (DiaPro Allegro/Largo 9 $\mu$m and DiaPro Dac 3 $\mu$m, Struers, Copenhagen, Denmark) for 3~minutes and 2 minutes per side, respectively, while maintaining the same monitoring pauses every 30 seconds.
And 3) finishing the process with oxide polishing suspensions (OP-S, Struers, Copenhagen, Denmark; Plate: MD-Chem, Struers, Copenhagen, Denmark; speed: 150~rpm; applied pressure: 10~N) for 1 minute.

Ultra-thin microtome slices of the tooth, 250~nm thick, were prepared using the same procedure as for bamboo samples for transmission spectroscopy, with the sliced tooth mounted in epoxy resin. Additionally, the original cross-sections used for ultra-thin microtome processing were investigated in reflection mode using a tabletop Bruker Vertex 70 FTIR spectrometer coupled with a Hyperion 1000/2000 FTIR microscope equipped with a wire-grid polarizer (Bruker Optik GmbH, Ettlingen, Germany).

\subsection{Synchrotron-FTIR microspectroscopy}

Synchrotron-FTIR measurements were conducted on the Infrared Microspectroscopy (IRM) beamline at the Australian Synchrotron (Victoria, Australia), using a Bruker Vertex 80v spectrometer coupled with a Hyperion 3000 FTIR microscope and a liquid nitrogen-cooled mercury cadmium telluride (MCT) detector (Bruker Optik GmbH, Ettlingen, Germany). 

The FTIR mapping measurements were performed in transmission mode using a matching pair of 36$\times$ IR objective and condenser (numerical aperture $NA = 0.50$; Bruker Optik GmbH, Ettlingen, Germany). 
An automate KRS-5 IR wire grid polarizer (Pike Technologies, Madison, WI, USA) was positioned in the path of the incident IR beam prior to the sample, and four spectral maps were acquired from the same region of sample at polarization angles of 0$^\circ$, 45$^\circ$, 90$^\circ$ and 135$^\circ$. All synchrotron-FTIR spectra were collected within a spectral range of 3900‒750~cm$^{-1}$ using 4 cm$^{-1}$ spectral resolution and 8 co-added scans. Blackman-Harris 3-Term apodization, Mertz phase correction, and zero-filling factor of 2 were set as default acquisition parameters using OPUS 8 software suite (Bruker Optik GmbH, Ettlingen, Germany). At each polarization angle, a new background spectrum was collected from a clean area on the same IR window using 64 co-added scans.

\begin{figure*}[tb]
\centering\includegraphics[width=18cm]{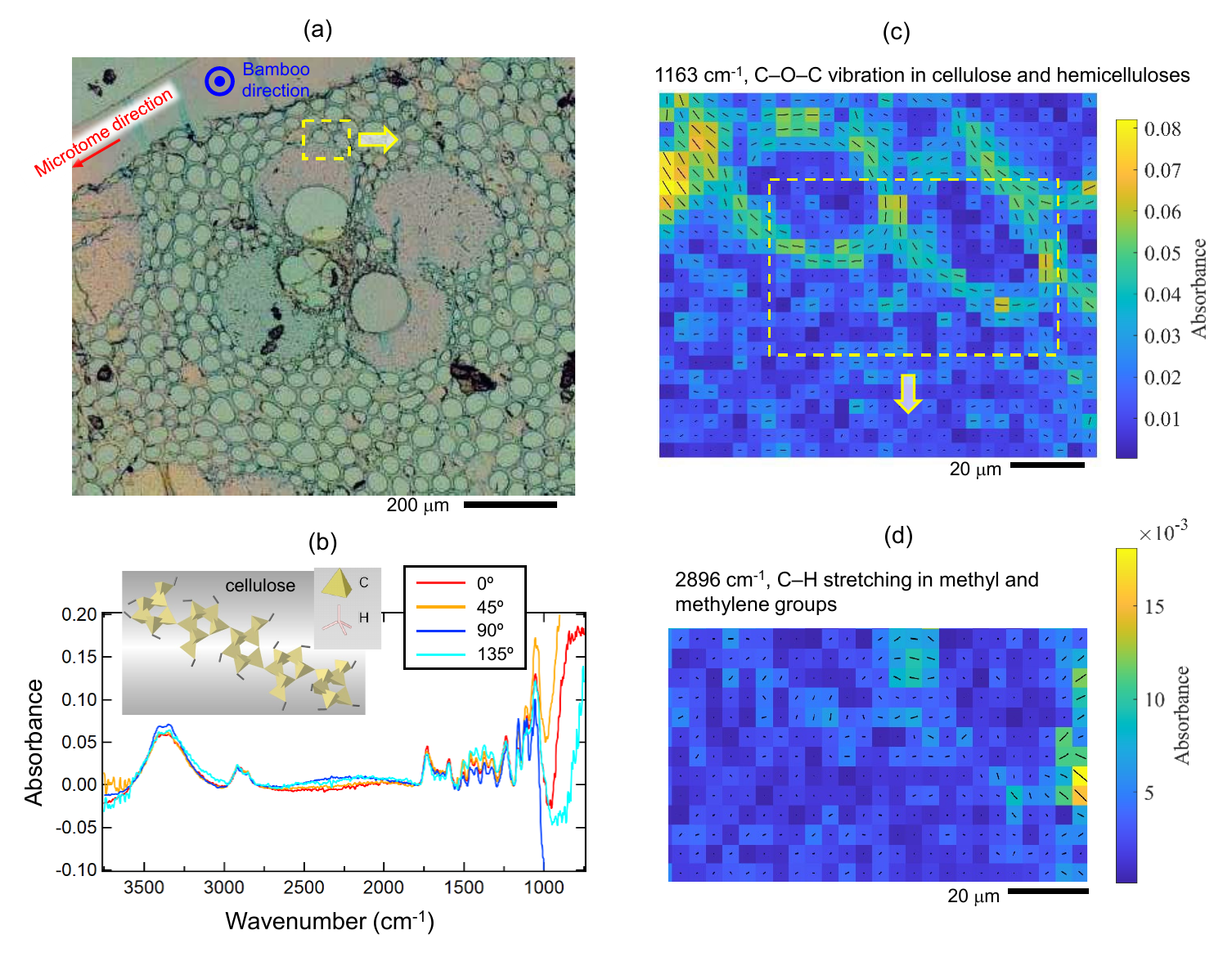}
  \caption{(a) Optical image of a microtome bamboo slice  on a \ce{Si3N4} 100-nm-thick membrane. (b) Single pixel IR absorbance spectra (a $4~\mu$m pixel corresponds to the step size in point-by-point scanning). The inset shows the cellulose structure, which is the same as in the inset of Fig.~\ref{f-lsli}(b), only in a polyhedral presentation. (c) Absorbance map at 1163~cm$^{-1}$ C-O-C band. Single pixel $4~\mu$m is larger in size as compared with the wall region between adjacent cells (see the high resolution image in Fig.~\ref{f-hires}). (d) Absorbance map at 2896~cm$^{-1}$ C-H band.} \label{f-tsli}
\end{figure*}
\begin{figure*}[tb]
\centering\includegraphics[width=18cm]{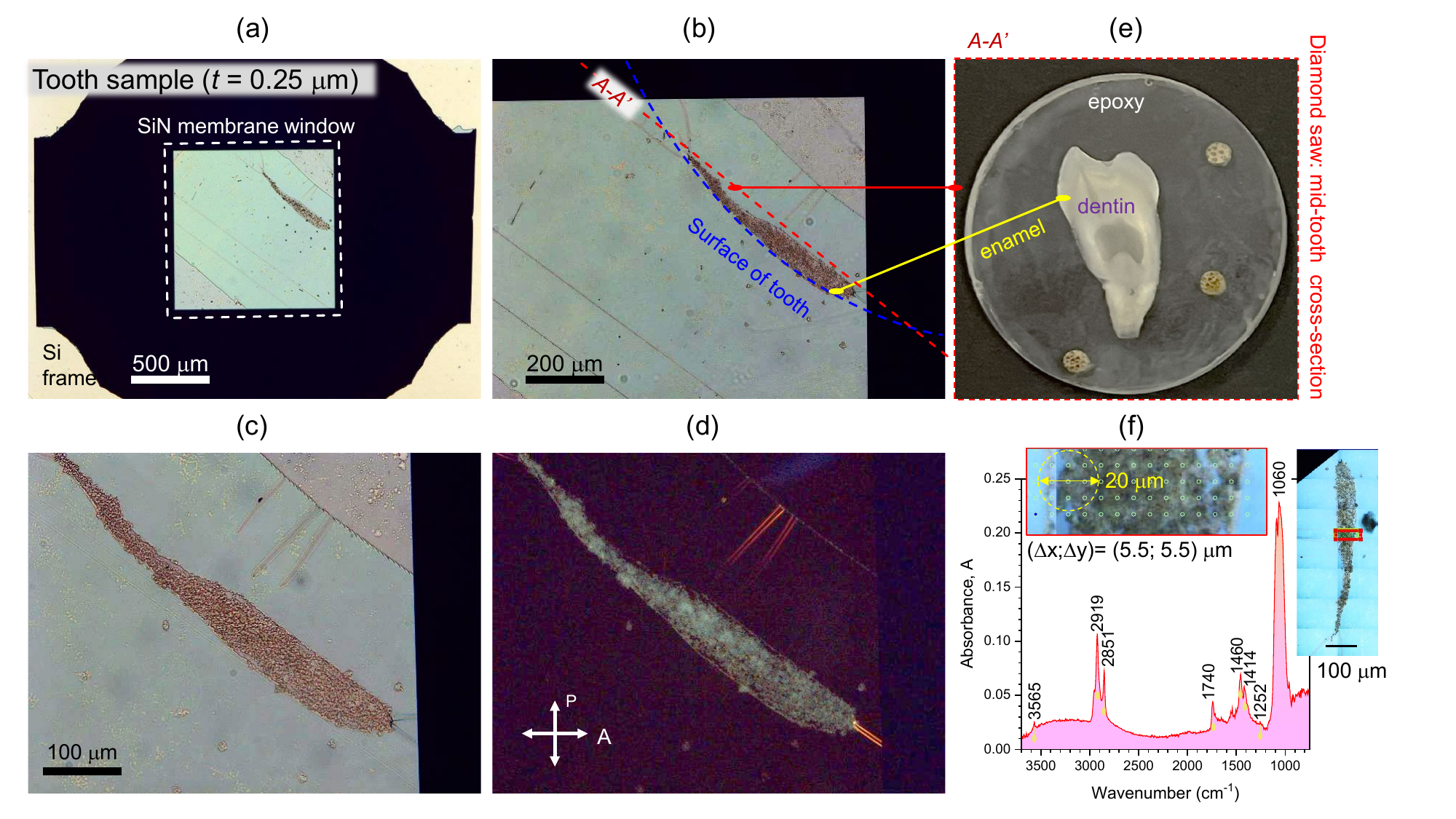
}
  \caption{Microtome of a tooth. (a) A 250-nm-thick micro-tome slice on a 100-nm-thick \ce{Si3N4} window for IR transmission. (b, c) Close up views of the slice. (d) Cross-polarized image of a micro-tome cross-section. (e) The first step of tooth preparation involves cutting through the middle by; 1) placing it into an epoxy puck, 2) cutting through with a diamond saw, and 3) polishing it, as described in Sec.\ref{MTooth}. This slice was used for the next micro-tome cut shown in (a-d). (f) Selected single point absorbance $A = -\lg T$ IR spectra using unpolarized synchrotron radiation at the enamel region; insets show optical images of the slice and matrix points separated by 5.5~$\mu$m. Tooth No.1; ethics agreement number 2022/9-1458-93.} \label{f-tooth}
\end{figure*}
\begin{figure}[tb]
\centering\includegraphics[width=8.5cm]{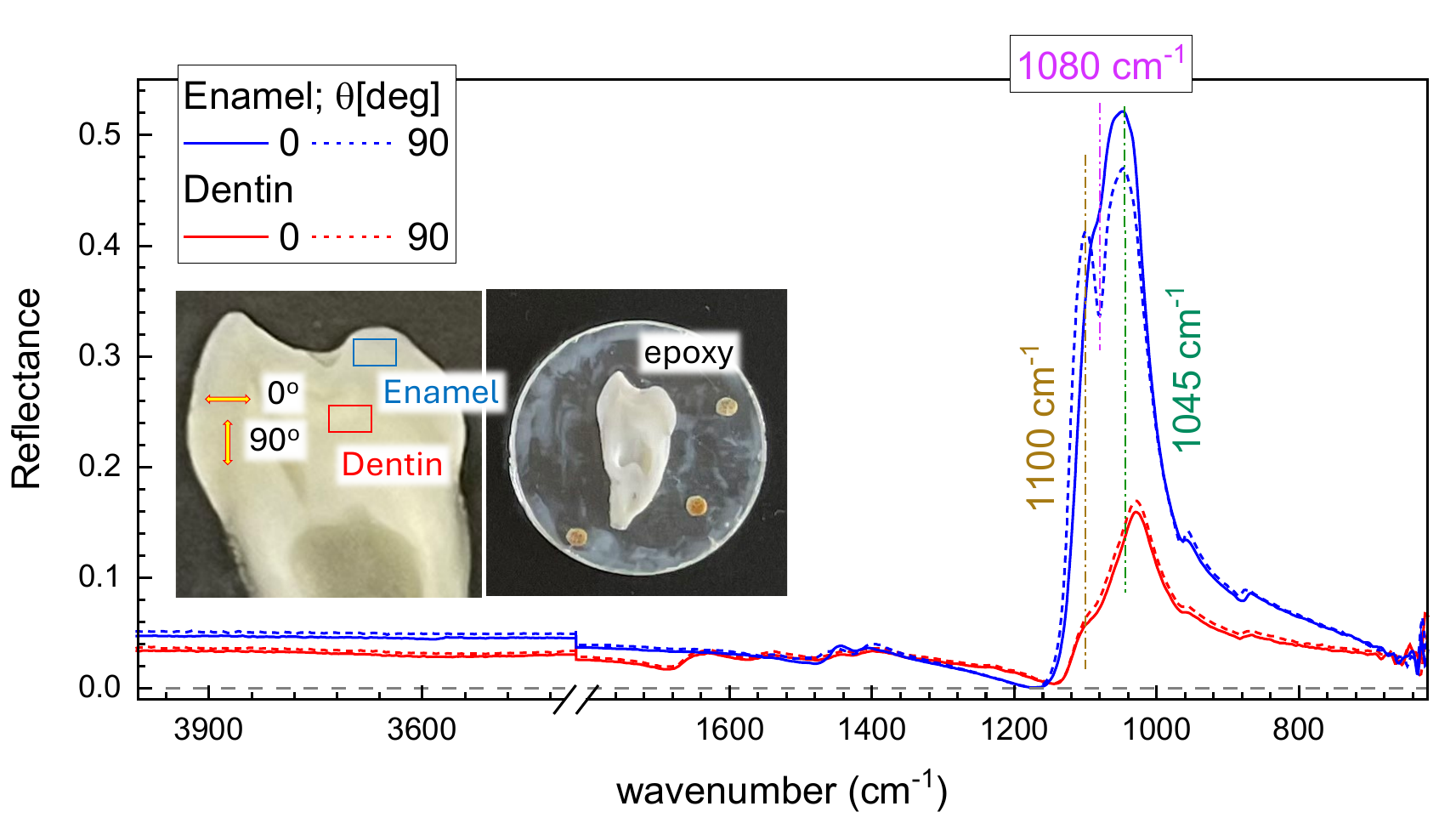}
  \caption{Reflectance spectra of enamel and dentin from a microtome slice of the tooth were measured at a single site. The numerical aperture of the objective lens was $NA = 0.4$. The inset shows the actual sample and the region of measurements. The tooth sample used here was No.~1 (patient younger than 18 years old).} \label{f-ftir}
\end{figure}

\subsection{The 4-pol. method}

The data analysis is based on detecting transmission spectra at four linear polarization angles, which differ from each other by $45^\circ$~\cite{4}. A non-linear least squares method is used to fit the experimental data to a function that models the relationship between absorptance ($\alpha$) and polarization angle ($\gamma$; Eq.~\ref{eq1}):~\cite{4,25arxiv}
\begin{equation}\label{eq1}
\alpha_\gamma = A_0\sin 2\gamma + A_1\cos 2\gamma +A_2,
\end{equation}
\noindent here, $A_0$, $A_1$, and $A_2$ are the fitting parameters obtained from the curve fitting. The in-plane azimuth angle of the transition dipole moment (TDM) $\Psi$, the dichroic ratios $D_{max}$ or $D_{min}$, and Herman’s orientation function $f_\Psi\in [0,1]$, where 0 - disorder and 1 ordered/aligned~\cite{4}: 
\begin{equation}\label{eq2}
    \Psi = \frac{1}{2}\tan^{-1}\left(\frac{A_0}{A_1}\right),
\end{equation}
\begin{equation}\label{eq3}
    D_{max} = \frac{2A_2 +2\sqrt{A_0^2 + A_1^2}}{2A_2 -2\sqrt{A_0^2 + A_1^2}},
\end{equation}
\begin{equation}\label{eq4}
     D_{min} = \frac{2A_2 -2\sqrt{A_0^2 + A_1^2}}{2A_2 +2\sqrt{A_0^2 + A_1^2}},
\end{equation}
\begin{equation}\label{eq5}
    f_\Psi = \frac{D_{max}-1}{D_{max}+2}\times\frac{2}{3\cos^2\beta -1},
\end{equation}
\begin{equation}\label{eq6}
    f_\Psi = \frac{D_{min}-1}{D_{min}+2}\times\frac{2}{3\cos^2\beta -1}.
\end{equation}


The azimuth angle $\Psi$ is defined as the angle at which maximum absorption occurs~\cite{4}. 
Calculation of the orientation function through the combination of Eqs.~\ref{eq3} and ~\ref{eq5}, or Eqs.~\ref{eq4} and ~\ref{eq6} is dependent on the relative orientation of the TDM with respect to the axis of the molecular chain, 
$\beta$. For vibrational modes that exhibit parallel dichroism, $\beta<54.73^\circ$, Eqs.~\ref{eq3} and~\ref{eq5} are used. Conversely, Eqs.~\ref{eq4} and ~\ref{eq6} are used for perpendicular dichroic vibrational modes, $\beta > 54.73^\circ$. 
The amplitude, $\alpha_{max}-\alpha_{min}=2\sqrt{A_0^2+A_1^2}$, and the approximate isotropic intensity, $(\alpha_{max}+\alpha_{min})/2=A_2$, are calculated based on the fitting parameters. 

\section{Results}

\subsection{Bamboo}

Assignment of absorption bands of cellulose is as follows (Fig.~\ref{f-poin}(a, b)): (1) 2896~cm$^{-1}$ C-H stretching in methyl and methylene groups (perpendicular dichroism), (2) 1425~cm$^{-1}$, \ce{CH2} deformation in cellulose, (3)~1370~cm$^{-1}$, C-H deformation in cellulose and hemicellulose, (4) 1163~cm$^{-1}$ C-O-C vibration in cellulose and hemicellulose (parallel dichroism), (5) 1110~cm$^{-1}$ C-O stretching and ring asymmetric valence vibration~\cite{bands}. Figure~\ref{f-poin} shows spectra at shorter and longer IR windows in (a) and (b), respectively measured in transmission from a single pixel of the map (c). The map of the longitudinal slice at the C-O band of 1110~cm$^{-1}$ shows a pattern of diagonal structures with stronger absorbance. Optical images of the longitudinal slice are shown in Fig.~\ref{f-lsli}(a) with a strong signature of birefringence (bright regions at diagonal orientation in cross-polarized image). We selected a region where parenchyma cells and vascular bundles are present (b). The 4-pol. measurements were carried out at C-O-C and C-N bands shown in (c) and (d), respectively. Those two vibration modes are known to be perpendicular (see inset in Fig.~\ref{f-poin}(c)). Indeed, a perpendicular orientation for the strongest absorbance was confirmed. The length of the line in each pixel represents the amplitude $(\alpha_{max} - \alpha_{min})$ and orientation shows the $\Psi$ angle/orientation. 

From the longitudinal slice, it was observed that there is no dichroism in cellulose absorption at the parenchyma cell wall, while there is high dichroism in cellulose at the vascular bundles (Fig.~\ref{f-lsli}). From the transverse slice, it was observed that there is high dichroism in cellulose absorption at the parenchyma cell wall and no dichroism in cellulose at vascular bundles (Fig.~\ref{f-tsli}); this is reverse to the case analyzed in the longitudinal slice above. Transverse slice imaging revealed that the C-O-C cellulose vibration clearly highlights the orientation of the walls, recognizable as elliptical structures formed by the alignment of the $\Psi$ angle with a large amplitude. It is formidable that pixel tracing of cells' walls can be achieved considering sub-wavelength thickness of $0.3~\mu$m slice, $8.6~\mu$m of the vibration band, and $4~\mu$m pixel size discussed in next section~\ref{disco}. The hyper-spectral imaging using an FTIR spectrometer facilitates vast possibilities for image analysis after data acquisition at a chosen spectral band.

\subsection{Tooth}

Figure~\ref{f-tooth} shows a 250~nm ultra-thin microtome slice of the tooth enamel region, which was used for transmission measurement. Variations of transmittance spectra from neighboring regions separated by $\sim 5.5~\mu$m were significant. This forbids a reliable polarization analysis in transmission. Figure~\ref{f-tooth}(f) shows representative single-point absorbance $A$ spectra with several characteristic peaks marked, which are close to the established bands (the assignment follows ref.~\cite{Rai,Liu_2014a,Orsini_2021}): 
C-H stretching of Amide B at 2920~cm$^{-1}$ and 2857~cm$^{-1}$, 
Amide A (\ce{N-H}) and \ce{OH} stretching at 3300~cm$^{-1}$ , 
Amide I (\ce{C=O}) at $\sim$ 1655~cm$^{-1}$, 
Amide II  (\ce{N-H}) at $\sim$ 1242~cm$^{-1}$, 
Amide III at $\sim$ 1235~cm$^{-1}$. 
The peaks 1412~cm$^{-1}$ and 1457~cm$^{-1}$ were contributed by type B $\nu_3$ mode carbonate \ce{CO3$^{2-}$},
while type B carbonate $\nu_1$ mode is at around 1069~cm$^{-1}$. 
The phosphate (\ce{PO4$^{3-}$}) $\nu_3$ vibration at $\sim$1069~cm$^{-1}$, $\sim$1052~cm$^{-1}$, and $\sim$1043~cm$^{-1}$~\cite{Rai,Liu_2014a,Orsini_2021} are prominent in Fig.~\ref{f-tooth}. This shows that ultra-thin slices, 10-to-50 times thinner than the wavelength, can be measured in transmission using a synchrotron radiation source. 

The other slice from the same tooth was measured in reflection at two polarization orientations of the incident beam, at $\theta = 0^\circ$ and $\theta = 90^\circ$, to reveal anisotropy (Fig.~\ref{f-ftir}). The strongest reflectance $R$ changes were at 1080~cm$^{-1}$ and 1100~cm$^{-1}$ for the spectral band shape and at 1045~cm$^{-1}$ for the amplitude at the edge enamel region of tooth's cross-section. According to established band assignments~\cite{Rai}, the typical strong absorbance $A$ bands (reduced transmittance $T$) are at the characteristic phosphates band at 1035~cm$^{-1}$ due to the $\nu_3$ asymmetric stretching of \ce{PO4$^{3-}$}. It is strong at the enamel region and shows anisotropy $\frac{R(0^\circ)}{R(90^\circ)}>1$. For the enamel, the B carbonate \ce{CO3$^{2-}$} $\nu_1$ mode at around 1069~cm$^{-1}$ is recognizable by a change of spectra band shape with a dip in $R$ as well as the type A carbonate band $\nu_1$ at 1104 cm$^{-1}$~\cite{Rai}. This spectral region of 1200 - 900~cm$^{-1}$ had no anisotropy in $R$ for the dentin section (Fig.~\ref{f-ftir}).

The spectral bands of the primary amides: amide I 1700 - 1600 cm$^{-1}$, amide II 1600 - 1500~cm$^{-1}$, amide A 3400 - 3350~cm$^{-1}$, and amide B 3085 - 3070 cm$^{-1}$ were barely distinguishable in both enamel and dentin $R$ spectra (Fig.~\ref{f-ftir}). 
Those spectral bands are less expressed as compared with the transmittance results, even from a very thin sample. The reflectance spectra are expected to be slightly different as compared with absorbance since $R$ has a contribution from $n$ and $\kappa$ (the refractive index $(n+i\kappa)$), where only $\kappa$ is related to $A$; the absorption coefficient is $\alpha=\frac{4\pi\kappa}{\lambda}$~[cm$^{-1}$].

\begin{figure*}[tb]
\centering\includegraphics[width=18cm]{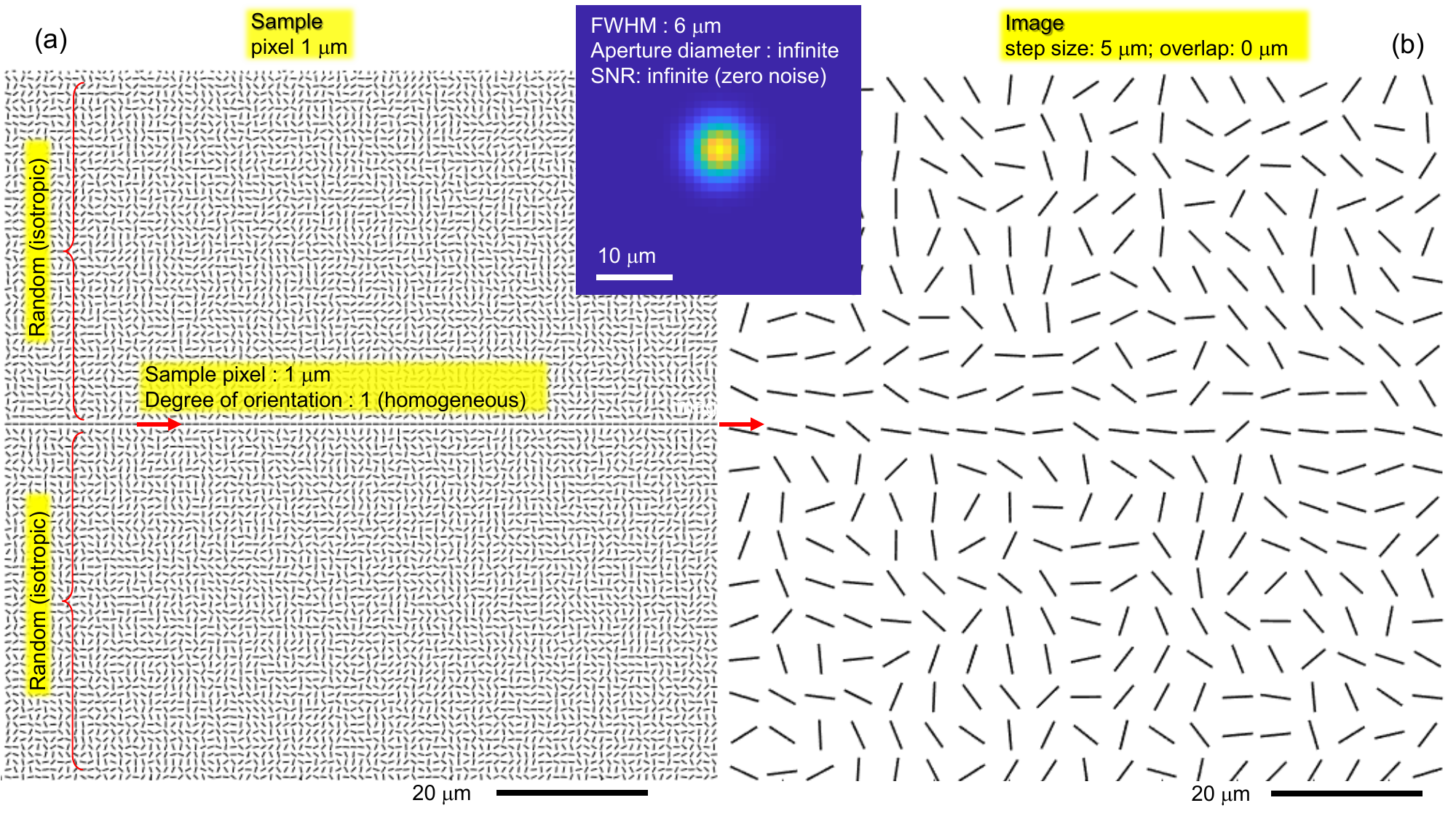}
  \caption{Simplified-model: (a) sample with 1~$\mu$m pixels with random orientation except the center horizontal line with all horizontally aligned pixels and (b) the image of orientation made by 4-pol. data analysis using Gaussian beam with 6~$\mu$m (FWHM) beam size, corresponding to the (10.2~$\mu$m diameter at $1/e^2$-intensity level); the inset shows the Gaussian beam.} \label{f-toy}
\end{figure*}

\section{Discussion}\label{disco}

\subsection{Resolution using 4-pol. method}

When an ultra-thin slice is used for imaging, a high resolution is obtained at specific absorption bands, as shown here. In a related problem of 3D imaging, the use of a confocal aperture or a pinhole before the detector was proposed to obtain a higher resolution by suppressing out-of-focus information~\cite{Franco}.

A noticeable enhancement of resolution is observed in IR absorbance mapping (Fig.~\ref{f-lsli}(c, d)), which needs further investigation. Chemical absorbance maps were taken with $NA = 0.5$ numerical aperture objective lens. For the $\lambda = 10~\mu$m (wavenumber $\tilde{\nu} = 1000$~cm$^{-1}$), the diffraction limit is $\lambda/NA\approx 20~\mu$m (see circle marker in Fig.~\ref{f-lsli}(c)). During data acquisition via step-scan in the xy-plane with a step of $\Delta_{x,y} = 4~\mu$m (which is the pixel size), there was an accumulation from approximately $5\times 5$ pixels from the focal spot in the xy-plane to each central pixel. This averaging process contributed to an improved S/N ratio. More importantly, it contributed to the high-resolution (mainly defined by the pixel size) of the absorption anisotropy image (map), which is due to the molecular alignment and structure in the specimen (a microtome slice of bamboo in this study). Under a linearly polarized incident beam with electrical $\mathbf{E}_{in}$-field, the absorbed portion of the light is defined by the interaction energy $\epsilon$ governed by the angular alignment of the absorbing dipole $\mathbf{p}$ and light field $\mathbf{E}_{in}$ in the sample plane xy. The interaction energy is given by $\epsilon = -(\mathbf{E}_{in}\cdot\mathbf{p})$.

Electrical dipole produces a field which scales as $E_d\propto\frac{1}{r^2}$, hence decays faster with distance as a field from a fixed charge $\propto\frac{1}{r}$. At the detector, where an optical system of a microscope delivers all light fields from the sample plane, there will be different dipole contributions from the on-axis (the central pixel) and from the side of the Gaussian focal spot of $\lambda/NA\approx 20~\mu$m (which is $\sim 5\times 5$ pixel matrix). Hence, the central pixel has a significant contribution due to its shortest distance and the highest intensity (at the center of the Gaussian-like intensity dome). The outer pixels have an increasingly lower contribution to the detected light due to lower intensity on the wings of Gaussian focus as well as larger distance. Moreover, the pixel-to-pixel distance is comparable with wavelength $\Delta_{x,y} \approx \lambda = 4~\mu$m as well as the thickness of the slice $t\ll\lambda$. This makes interference a major contributor to the measured signal at the center pixel. As the sample is scanned in steps of 5~$\mu$m, an averaging takes place. However, the center pixel has always the dominant contribution to the absorption map at each step. 

Moreover, the cell walls are strongly birefringent in the visible spectral region (Fig.~\ref{f-hires}), and birefringence is expected in the IR spectral range due to the gradient of refractive index $\Delta n$. This causes a stronger reflectance $R$, which contributes to a lower transmittance $T$, hence a larger absorbance $A$ ($A+R+T=1$). Clearly, the cell walls are thinner than the step/pixel size of $\sim 4~\mu$m in the image (Fig.~\ref{f-hires}) and are below the diffraction limit of $\lambda/NA\approx 20~\mu$m under the used imaging conditions. It is apparent that only the cell walls have a strong orientation (Fig.~\ref{f-tsli}(c)) and are resolved with the used pixel/step size of $4~\mu$m. This qualitatively explains why the absorption anisotropy can be revealed at the sub-diffraction limit. Here, we report on experimental observation, which will need further theoretical analysis. A simple model is presented below for numerical validation of the high resolution (Sec.~\ref{super}).

\subsection{Simplified model for numerical validation}\label{super}

To numerically test the possibility of detecting a pixel with a fixed orientation of absorbance in the otherwise randomly oriented matrix, a simplified model to test 4-pol. method was made (Fig.~\ref{f-toy}). The Gaussian beam was defined with the possibility of adding an aperture and random noise. The beam size, up to a few times larger compared with the pixel size, was scanned over the matrix and 4-pol. orientation was determined to simulate the actual experiments. The pixel was set $1~\mu$m similar to the cell boundary in the actual bamboo cell (Fig.~\ref{f-hires}). Only the center line was set with a fixed orientation, i.e. the orientation of absorbers was fixed horizontally for the entire center line, while the rest of the pixels had a random orientation. The Gaussian beam at the $1/e^2$-intensity level was approximately ten times larger. Imaging was carried out with $\Delta x = \Delta y = 5~\mu$m step. Figure~\ref{f-toy}(b) shows the orientation map with a recognizable center line, which has a preferential horizontal orientation. No aperture nor random noise was added on the readout Gaussian profile. This type of enhanced resolution has a similarity with lock-in detection. A digital demodulation of a single oriented pixel in an array of many non-oriented random pixels is detected by its $\pi$ orientation folding (period) using the 4-pol. method. Due to this principle, the detection limit (maximum S/N ratio) for the isolated structure may be larger than that of usual imaging based on intensity distribution.

\subsection{Ultra-thin microtome of tooth}

Recent studies of teeth structure revealed that microcracks are an integral part of the whole healthy tooth~\cite{21sr14810,dumi2022}. Characterization of microcracks with a micrometer-wide scale is carried out by UV light excited photoluminescence~\cite{dumi2025}, X-ray micro-computed tomography in a non-destructive way, i.e without cutting the tooth
~\cite{23b1354}. 

This study shows that ultra-thin sub-$\mu$m slices of the tooth can be prepared and used for transmission and reflection FTIR spectroscopy for the chemical fingerprinting spectral range of 3-15~$\mu$m. Such cross-sections can open the next level of details at a higher resolution for understanding the effects of form-birefringence due to tubular enamel structure, as well as the (re/de)mineralization of the crystalline hydroxyapatite \ce{Ca10(PO4)6(OH)2} in the enamel~\cite{inpress}.

\section{Conclusion and outlook}

It is shown that sub-wavelength $\sim 0.3-0.5~\mu$m thickness of bamboo slices made in the longitudinal and transverse directions complement each other in spectral analysis using 4-pol. method for the determination of molecular alignment in the cellulose building blocks of parenchyma cells and vascular bundles in bamboo. Intricate orientation patterns can be revealed at the selected IR bands of the absorbance spectra. 

Ultra-thin microtome slices of even brittle composite materials such as enamel and dentin in teeth can be made. It broadens a toolbox for material characterization at the particular regions of biological structures, such as teeth microcracks (width range, 0.25 - 35.00~$\mu$m; length range, 0.24 - 10.00~$\mu$m)~\cite{21sr14810,dumi2022}, non-caries lesions, interfaces of dental fillings and restorations, etc., which are of great interest and importance to medical society and bio-sciences.

The transmission measurements are usually preferred as compared to the reflectance due to the known probed volume and the possibility of quantitative determination of absorbance and its dichroism. Microtome is an essential sample preparation tool. Characterization of sub-wavelength thickness samples is usually challenging, however, using the 4-pol. method, information on absorption dichroism is revealed, as shown in this study. The absorbance anisotropy can be also measured using non-propagating near-field light in the ATR geometry as was recently demonstrated for the non-imaging mode~\cite{22nh1047}. Since the ATR technique is widely used in biomedical research, it would be very motivating to develop a multi-spectral imaging version to reveal anisotropy and alignment in the IR spectral range since they are linked to the development of some of the most devastating medical conditions. It is shown by the numerical model that the orientation anisotropy (with dimensions ranging from 25~nm up to $\approx$ 100 nm) can be revealed even below the spatial resolution, which is valuable in the IR spectral range and offers non-destructive biomedical imaging.

\bibliography{apssamp}
\begin{acknowledgments}
The polarization synchrotron-FTIR analysis was undertaken at the Infrared Microspectroscopy (IRM) beamline at the Australian Synchrotron, part of ANSTO, via merit-based beamtime proposal (ID. M22505) in 2024. The 4-pol. method was made available for the IRM users from 2019. SJ is grateful for support via the Australian Research Council Discovery DP240103231 grant. This work was supported by JST CREST Grant Number JPMJCR19I3, Japan. MM acknowledges the framework of the "Universities` Excellence Initiative” programme by the Ministry of Education, Science and Sports of the Republic of Lithuania under the agreement with the Research Council of Lithuania (project No. S-A-UEI-23-6).
\end{acknowledgments}

\setcounter{figure}{0}\setcounter{equation}{0}
\setcounter{section}{0}\setcounter{equation}{0}
\makeatletter 
\renewcommand{\thefigure}{A\arabic{figure}}
\renewcommand{\theequation}{A\arabic{equation}}
\renewcommand{\thesection}{A\arabic{section}}
\appendix
\section{Appendix}

\begin{figure*}[h!]
\centering\includegraphics[width=18cm]{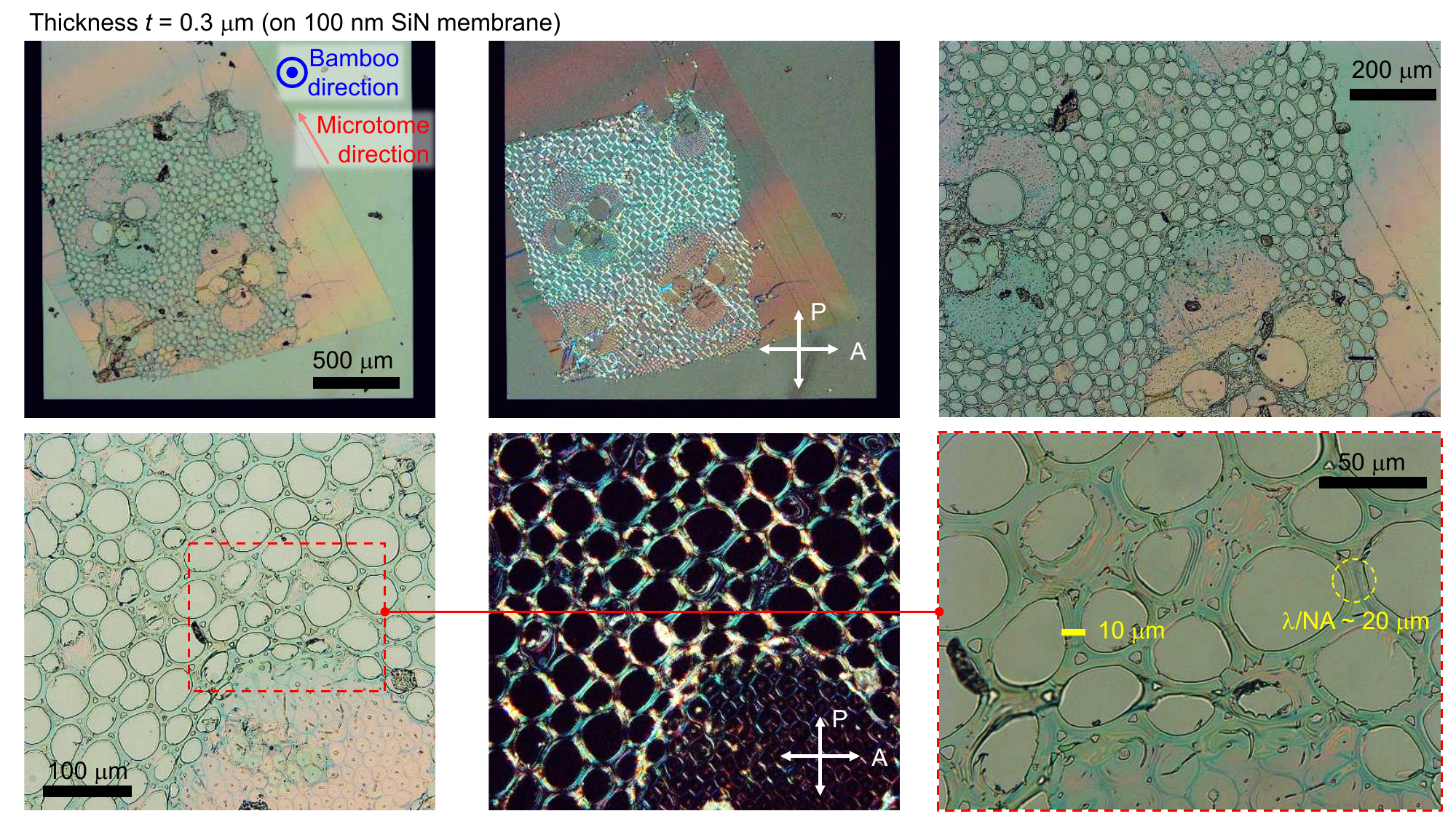}
  \caption{Optical images of the bamboo ultra-thin microtome (Leica EM UC6) sub-wavelength transverse slice at different resolutions and using the cross-polarized mode. Bamboo sample was taken from an ordinary chopstick. The diffraction limit at the used conditions for IR imaging with synchrotron radiation at $NA=0.5$ and wavelengths around $\lambda = 10~\mu$m was $\sim 20~\mu$m (see the dashed-line circle marker).} \label{f-hires}
\end{figure*}

\end{document}